      [1999/12/01 v1.4c Il Nuovo Cimento]
\documentclass{cimento}
\usepackage{psfig, wrapfig}

\title{Dust scattering X-ray expanding rings around gamma-ray bursts}
\author{S.~Mereghetti\from{ins:x},
A.~Tiengo\from{ins:x}, G.~Vianello\from{ins:y}
}

\instlist{\inst{ins:x} INAF - IASF Milano, via Bassini 15, Milano
I-20133,       Italy
  \inst{ins:y} Universit\'{a} degli Studi di Milano, Dip. di Fisica,
  v. Celoria 16, Milano  I-20133, Italy}
\PACSes{\PACSit{???}{???}
}
\begin{document}

\maketitle

\vspace{-6cm}
\begin{center}
 \textbf{Proceedings of the international conference
\textsl{Swift and GRBs: Unveiling the Relativistic Universe} -
Venice, 5-9 June 2006}
\end{center}
\vspace{4cm}

\begin{abstract}
Scattering by dust grains in our Galaxy can produce X--ray halos,
visible as expanding rings,  around GRBs. This has been observed
in three GRBs to date, allowing to derive accurate distances for
the dust clouds as well as some constraints on the prompt GRB
X--ray emission that was not directly observed. We developed a new
analysis method to study dust scattering expanding rings and have
applied it to all the XMM-Newton and Swift/XRT follow-up
observations of GRBs.
\end{abstract}

\section{Introduction}

Soon after the discovery of celestial X--ray sources it was
realized that scattering from interstellar dust grains could lead
to the formation of detectable X--ray halos surrounding the source
images~\cite{over}. The usefulness of this phenomenon to study the
properties of the dust was pointed out by several authors, but
observations had to await the development of imaging X--ray
telescopes. The first dust halos  were discovered with the
Einstein Observatory around bright galactic sources at the
beginning of the eighties~\cite{rolf,catura}. Currently, the
modelling of the energy-dependent radial profiles of X--ray halos
is a well established tool in the study of the interstellar dust.

Owing to the longer path lengths of scattered compared to
unscattered photons, variability in the source can lead to
time-dependent changes in the halo radial profile. If the spatial
distribution of the dust along the line of sight is known, it is
possible to constrain the distance of variable sources by studying
their halos~\cite{ts73}. This method was also
proposed~\cite{klose} as a way to distinguish between the galactic
and cosmological origin of GRBs, but it could not be applied due
to the lack of sensitive imaging detectors.

Only recently dust halos have been detected around three bursts:
GRB 031203~\cite{vaughan04},  GRB 050724~\cite{vaughan06} and GRB
050713A~\cite{tm}. Due to the short duration of the bursts and the
relatively small thickness of the dust layers, such halos appear
as expanding rings. Since the scattering dust is in our Galaxy, at
a distance d$_s$ much smaller than that of the GRB, the ring
radius $\Theta$ and the time delay $\Delta$t = t--t$_{GRB}$ are
simply related by
 $\Delta$t = (d$_s$/2c)~$\Theta^2$.
Thus it is possible to accurately measure the dust distance d$_s$
by fitting the time expansion of the ring.

The brightness of the ring depends on the intensity of the GRB
emission I$_{GRB}$ and on the scattering optical depth $\tau$ :
I$_{HALO}$ = I$_{GRB}$ (1--e$^{-\tau}$) $\simeq$ $\tau$ I$_{GRB}$.
If both I$_{HALO}$ and  I$_{GRB}$ are known, as in the case of GRB
050724 \cite{vaughan06}, some information on the amount of dust
and its properties can be derived. In the other two cases observed
to date this was not possible and the observations have instead
been used to constrain I$_{GRB}$ based on our best guesses on
$\tau$. The latter can in fact be estimated from the optical
extinction A$_V$ due to the dust layer, although the proper
relation between $\tau$ and A$_V$ is debated~\cite{PS95,DB04} and
also the measurement of A$_V$ can be affected by significant
uncertainties (in most cases only an estimate of the total A$_V$
along the line of sight is available). These problems are well
exemplified by the case of GRB 031203 discussed below.

\section{A new method for the detection of expanding rings in X-ray images}

\begin{figure}[h!]
\begin{center}
\hbox{ \hspace{-2.0cm}
 \psfig{figure=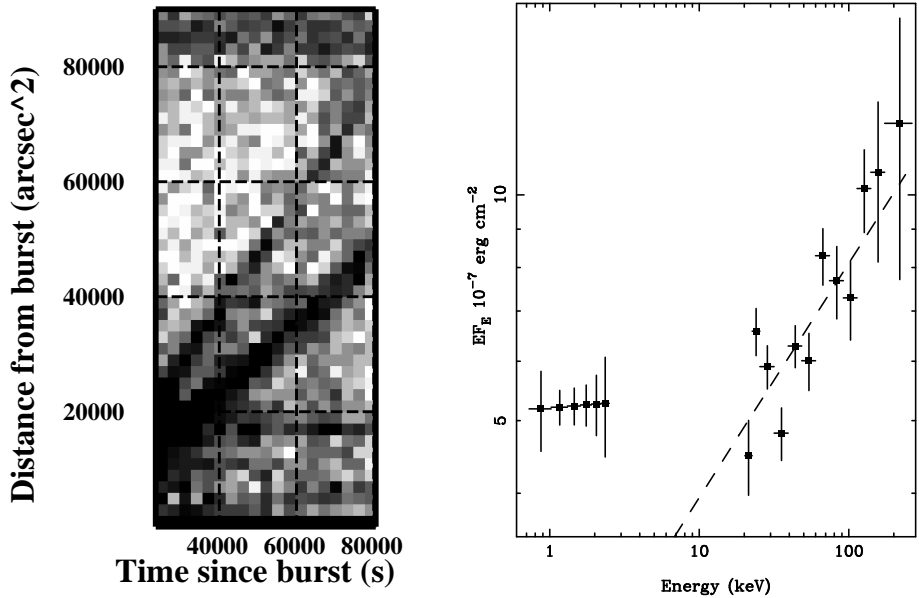,width=7.5cm,angle=0}
 \hspace{0.1cm}
\psfig{figure=mereghettis_fig1b.ps,width=6.0cm,angle=0} }
 \vspace{-1.0cm}
\end{center}
\caption{Left: Dynamical image of GRB 031203. The inclined lines
result from the concentric expanding rings due to two layers of
dust at distances of 870 pc (upper line) and 1384 pc. The
horizontal lines are X-ray point sources in the field. Right:
fluence spectrum of GRB 031203 as measured with IBIS/INTEGRAL
above 20 keV and reconstructed from the halo analysis below 3
keV.} \label{fig1}
\end{figure}

Bright dust scattering rings, like those seen around GRB 031203,
can be easily discovered and studied by comparing a sequence of
images obtained at different times. However, with this simple
method it is difficult to detect fainter halos, as demonstrated by
the case of GRB 050713A, whose halo was discovered thanks to a
different technique based on ``dynamical images" ~\cite{tm}. In a
dynamical image all the counts detected by the X--ray telescope
are binned according to their squared angular distance from the
GRB position (ordinate axis)  and their time from the GRB
(abscissa axis). Expanding rings are visible as inclined lines
whose slope is proportional to 1/d$_s$. The 1-2 keV dynamical
image for GRB 031203 based on EPIC/pn data is shown in
Fig.~\ref{fig1}.

 The dust distance and the flux in the ring can be
measured from the distribution of the quantity  D$_i$ $\equiv$ 2c
$\Delta$t$_i$/$\Theta_i^2$ which can be computed for each count.
Only for the scattered X--ray photons this quantity represents the
dust distance. Therefore, an expanding ring is visible as a peak
centered at d$_s$ in the distribution n(D) of the D$_i$ values. A
spatially uniform instrumental background gives a contribution
proportional to D$^{-2}$ in the range from D$_{min}$ = 2c
$\Delta$t$_{max}$/$\Theta^2_{max}$ to  D$_{max}$ = 2c
$\Delta$t$_{min}$/$\Theta^2_{min}$, where $\Delta$t$_{min}$,
$\Delta$t$_{max}$, $\Theta_{min}$ and $\Theta_{max}$ delimit the
rectangular region of the dynamical image from which the counts
are extracted.

\begin{figure}[ht!]
 \psfig{figure=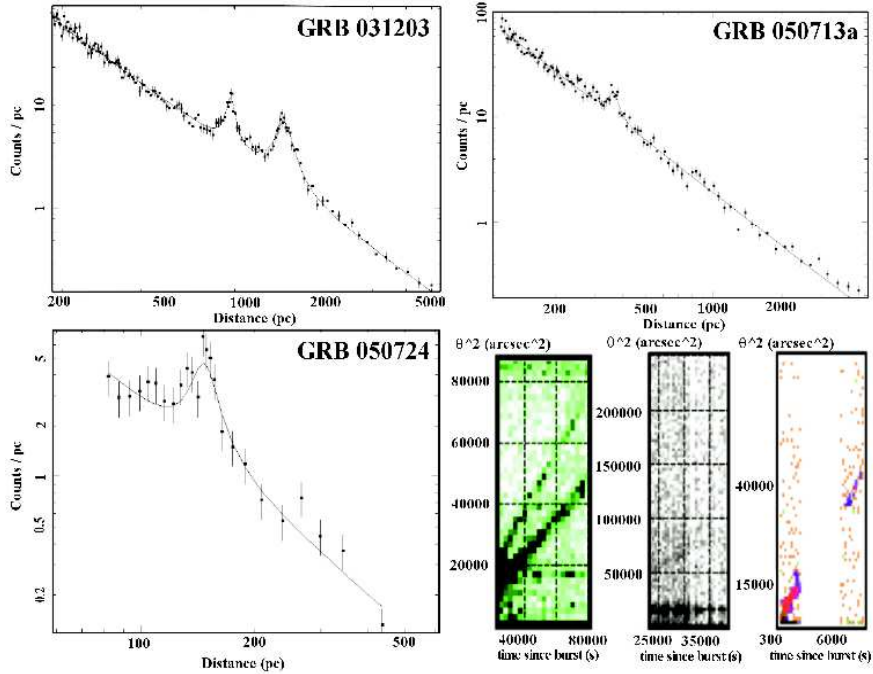,width=12cm,angle=0}
  \caption{Distributions  n(D) and best
fits with power laws plus Lorentzian curves for the three GRBs
with dust scattering expanding rings. The corresponding dynamical
images are shown in the bottom right panel.} \label{nd}
\end{figure}

The distributions of D$_i$  for the three GRBs are shown in
Fig.~\ref{nd}, where the peaks corresponding to the dust
scattering rings are clearly visible.  These distributions are
well fit with the sum of a power law with slope $\sim$--2,
representing the background contribution, and Lorentzian curves to
model the peaks. The net number of halo counts can then be
obtained by integrating the Lorentzians. If enough  counts are
present in the halos, this procedure can be done in different
energy bins to extract the halo spectrum. The results of our fits
for the three GRBs are reported in Table~\ref{tab1}.

\begin{table}
  \caption{GRBs with dust scattering halos}
  \label{tab1}
  \begin{tabular}{llll}
    \hline
                   &    GRB 031203   & GRB 050713A  & GRB 050724    \\
\hline
Galactic coordinates &     256, --5   & 112, +19    & 350, +15    \\
Galactic N$_H$   (10$^{20}$ cm$^{-2}$)   & 85 & 11 & 15    \\
 red-shift & 0.1055  & -- & 0.258 \\
 \hline
 GRB discovery        &    INTEGRAL/IBIS    & Swift/BAT  &      Swift/BAT    \\
Fluence (erg cm$^{-2}$) &    (1.66$\pm$0.08)$\times$10$^{-6}$ & (9.1$\pm$0.6)$\times$10$^{-6}$ & (6.3$\pm$1.0)$\times$10$^{-7}$  \\
                        &  (20--200 keV)  &  (15--350 keV) &    (15--350 keV)  \\
 Duration (s)         &  40  &  70 & 0.25 (spike) \\
                     &   &   &  + $\sim$200  (tail) \\
 \hline
Halo discovery       &    XMM-Newton/EPIC   & XMM-Newton/EPIC & Swift/XRT \\
Observed time interval (ks) &  23--80       & 23--46   &  0.34--2.2, 6.1--8.1   \\
dust layer A$_V$   &  2 & 0.5  & 1.5 \\
 d$_s$ (pc)       &  870$\pm$5, 1384$\pm$9 &  364$^{+6}_{-7}$  & 144$\pm$3 \\
I$_{HALO}$ (counts) & 840$^{+210}_{-180}$, $1740^{+270}_{-240}$   &  $185^{+120}_{-90}$  &   $155^{+167}_{-110}$ \\
%
%
\hline
  \end{tabular}
\end{table}

We applied this analysis method to all the GRBs observed to date
with XMM-Newton and Swift/XRT, without finding any other dust
ring. A comparison of the relative success rate of the two
satellites (2/15 for XMM-Newton versus 1/150 for Swift) indicates
that instruments with large collecting area are required to detect
such very faint structures.

\section{GRB 031203}

This nearby burst (z=0.105) is particulary interesting since it is
under-energetic, given its apparently normal hardness, to fit on
the correlations followed by most GRBs \cite{amati,ghirla}. Our
re-analysis of the INTEGRAL IBIS data with the most recent
software and calibrations confirms that the spectrum is well fit
by a power-law with photon index 1.69$\pm$0.06 (1$\sigma$). By
fitting with a Band spectrum, a 99\% c.l. lower limit of 100 keV
can be placed on E$_{peak}$. The 20-200 keV fluence over 40 s is
(1.66$\pm$0.08)$\times$10$^{-6}$ erg cm$^{-2}$. The fluence in
soft X-rays that we inferred from the dust scattering rings is
smaller than that derived in \cite{vaughan04}, but it still
exceeds the backward extrapolation of the INTEGRAL spectrum by a
significant factor (see Fig.\ref{fig1}, right panel). Thus it is
likely that the soft X--ray emission in GRB 031203 consisted of a
delayed component following the hard pulse seen with INTEGRAL,
similar to what has been observed in GRB 060218 \cite{ghis}. In
this case GRB 031203 would have a smaller E$_{peak}$ and a higher
luminosity, making it in agreement with the ``standard''
relations.


\end{document}